# Energy-Dependent Timing of Thermal Emission in Solar Flares

Rajmal Jain, Arun Kumar Awasthi, Arvind Singh Rajpurohit,

Physical Research Laboratory
(Dept. of Space, Govt. of India),
Navrangpura, Ahmedabad 380 009

and

Markus J. Aschwanden,

Lockheed Martin Advanced Technology Center,
Solar and Astrophysics Laboratory, California, USA

## Abstract

We report solar flare plasma to be multi-thermal in nature based on the theoretical model and study of the energy-dependent timing of thermal emission in ten M-class flares. We employ high-resolution X-ray spectra observed by the Si detector of the "Solar X-ray Spectrometer" (SOXS). The SOXS onboard the Indian GSAT-2 spacecraft was launched by the GSLV-D2 rocket on 8 May 2003. Firstly we model the spectral evolution of the X-ray line and continuum emission flux $F(\varepsilon)$ from the flare by integrating a series of isothermal plasma flux. We find that multi-temperature integrated flux $F(\varepsilon)$ is a power-law function of $\varepsilon$ with a spectral index ($\gamma$) ≈ -4.65. Next, based on spectral-temporal evolution of the flares we find that the emission in the energy range $E$= 4 - 15 keV is dominated by temperatures of $T$= 12 – 50 MK, while the multi-thermal power-law DEM index ($\delta$) varies in the range of -4.4 and -5.7. The temporal evolution of the X-ray flux $F(\varepsilon,t)$ assuming a multi-temperature plasma governed by thermal conduction cooling reveals that the temperature-dependent cooling time varies between 296 and 4640 s and the electron density ($n_e$) varies in the range of $n_e$= (1.77-29.3)*$10^{10}$ cm$^{-3}$. Employing temporal evolution technique in the current study as an alternative method for separating thermal from non-thermal components in the energy spectra, we measure the break-energy point ranging between 14 and 21±1.0 keV.

## 1. Introduction

The X-ray emission from solar flares has been traditionally treated as well as shown to be isothermal in nature (Landi *et al.,*2003, 2005, 2006; Phillips *et al.,*2003; Caspi and Lin, 2010), particularly while simulating spectral line and continuum intensities using CHIANTI atomic database (Landi and Klimchuk, 2010; Phillips *et al.,*2010). However, Craig and Brown (1976) formulated the concept of multi-temperature plasma in the flare considering differential emission measure (DEM). Aschwanden (2007) has for the first time investigated the nature of X-ray emission in solar flares with the help of modeling the spectral-temporal hard X-ray flux in terms of multi-temperature plasma governed by thermal conduction cooling. He suggested three physical processes that lead to measurable time delays as a function of energy in solar flares: 1) time-of-flight dispersion of free-streaming electrons, 2) collisional trapping of electrons, 3) and the Neupert effect (Hudson 1991; Dennis and Zarro 1993), *i.e.*, the thermal delays can be caused by the temperature dependence of the cooling process, such as thermal conduction, $\tau_c(T) \propto T^{-5/2}$ (*e.g.*, Antiochos and Sturrock 1978; Culhane *et al.,*1994) or radiative cooling $\tau_r(T) \propto T^{5/3}$. In his quantitative model, he has characterized the multi-temperature differential emission measure distribution (DEM) and non-thermal spectra with power-law functions. By fitting this model to the spectra and energy-dependent time delays on the RHESSI flare observations from 10 keV and above he suggested the flare X-ray emission to be multi-thermal in nature. However, the nature of the flare emission below 10 keV is still not addressed in the context of isothermal versus multi-thermal whose cause is mostly thermal bremsstrahlung. Thus it is extremely important to characterize the soft X-ray emission below 10 keV. Further, the X-ray emission between 10 and 25 keV should in principle include both thermal and non-thermal contributions (Aschwanden, 2007; Jain *et al.,* 2008). Therefore, the characterization of flare plasma up to 25 keV with high spectral resolution observations becomes further important.

Jain *et al.* (2008) showed that the X-ray bursts associated with solar flares above 25 keV come mostly from the non-thermal bremsstrahlung while the X-ray emission between 10 and 25 keV in principle includes both thermal and non-thermal contributions. Both types of X-ray emission occur during solar flares and the clear demarcation line between the two processes provides quantitative estimation of energy release by each process in solar flares. However, the estimation of the non-thermal energy content in the parent electrons giving rise to flare emission solely depends on the low-energy cutoff and/or the thermal-non-thermal crossover energy (break energy). However, so far no relation between the low energy cut-off of power-law electrons ($E_c$) and the break-energy point ($E_B$) has been established. Aschwanden (2007), ignoring possible low-energy cutoff, has found the thermal-non-thermal crossover energy 18±3.4 keV using the power law approximation of X-ray emission. On the other hand, Gan *et al.* (2002) estimated the low energy cut-off ($E_c$) of the non-thermal emission by fitting the hard X-ray spectra obtained from BATSE/CGRO. They found value of $E_c$ varies between 45 and 97 keV with an average of 60 keV for 44% of the events considered for investigation. For another 44% events they suggested that $E_c$ could be lower than 45 keV due to non-availability of data below 45 keV. On the other hand, for 11% of events the hard X-ray spectra could not be explained by a beam of power-law electrons with low-energy cutoff, which supports the break-energy concept and that could be on the lower side (Jain *et a.*, 2000; Aschwanden, 2007). However, Sui *et al.* (2005) found 24±2 keV as the low-energy cutoff ($E_c$) to ensure that always thermal emission dominates over non-thermal emission in low energy. They estimated the non-thermal energy content in the electrons of the order of 1.6 x $10^{30}$ ergs. On the other hand, Saint and Benz (2005) considering 20 keV as the turnover energy, which is perhaps the same as the break energy ($E_b$), estimated the non-thermal energy to be ≈2 * $10^{30}$ ergs, almost the same value as Sui *et al.* (2005) found for an M1.2 class flare. However, the low-energy cutoff seems physically not realistic as such a configuration leads to plasma instability. Such instabilities have a growth rate typically of the order of local plasma frequency, *i.e.*, orders of magnitude shorter than the propagation time of the beam within the acceleration region. Therefore, the turnover of break energy appears to be more physically realistic and needs to be measured as precisely as possible.

In the context of the above issues of isothermal versus multi-thermal and crucial need of measurement of break energy point we have undertaken the investigation of ten M-class flares observed by the Si detector of Solar X-ray Spectrometer (SOXS) experiment onboard the Indian GSAT-2 spacecraft.

## 2. Modeling the X-ray Flare

The bremsstrahlung spectrum $F(\varepsilon)$ of multi-thermal plasma with temperature $T$ as a function of photon energy $\varepsilon = h\nu$ is given by,

$$F(\varepsilon) = F_0 \int \frac{\exp(-\varepsilon/k_B T)}{T^{1/2}} \frac{dEM(T)}{dT} dT, \qquad (1)$$

where $F_0 = 8.1*10^{-39}$ keV s$^{-1}$ cm$^{-2}$ keV$^{-1}$ (Culhane, 1969; Craig and Brown, 1976) and dEM/dT specifies the temperature sensitivity of the differential emission measure in the volume element dV with temperature $T$ and electron density $n_e$ as,

$$\left[\frac{dEM(T)}{dT}\right] dT = n_e^2(T) dV. \qquad (2)$$

Jain *et al.* (2005) and Jain (2009) simulated the full-disk integrated photon spectrum $F(\varepsilon)$ from 0.01 keV to 1000 keV, considering a three-temperature emission measure (EM) distribution of the coronal plasma, including quiet (pre-flare) and various flare components of the M5 class flare. Based on the results from *Yohkoh* observations, they employed $T$ = 4MK, EM = $10^{49}$ cm$^{-3}$ for the pre-flare background, $T$ = 13 MK and EM = $10^{49.5}$ cm$^{-3}$ for the thermal component, and $T$ = 40 MK and EM = $10^{47}$ cm$^{-3}$ for the superhot component. They also considered a non-thermal component with a spectral index ($\gamma$) = -3.5 (whose range is between -2.5 and -4.5) and a flux of 10 photons cm$^{-2}$ s$^{-1}$keV$^{-1}$ at 20 keV.

In the current study we simulate the multi-thermal flux by integrating the series of isothermal photon flux considering line and continuum emissions and compare the observations in the energy range 4-15 keV in contrast to Aschwanden (2007). Aschwanden (2007) considered only continuum from 1 to 100 keV, but compared the observations of RHESSI only from 10 keV. In order to estimate the flux of isothermal plasma, we employ the temperature ($T$) and emission measure (EM) of the flare plasma from 'A.Awasthi private communication'. The details of estimating isothermal flux employing f_vth.pro in OSPEX of SolarSoft have also been described in 'A. Awasthi, private communication'. Shown in Figure 1 is the simulated line and continuum X-ray spectra for temperatures of $T$=11, 14,…34 MK in steps of 3 MK (thin curve) and their sum (dotted). The integrated flux is then fitted with the power-law function (the red line in Figure 1), which provides the photon spectral index ($\gamma$) of -4.65. However, if the flare plasma is multi-thermal then the photon spectral index of flux ($\gamma$) and power-law index of DEM ($\delta$) should be related as $\gamma = \delta - 0.5$ (Aschwanden, 2007; Klimchuk *et al.*, 2008). This suggests $\delta$ should be -5.15. Therefore in order to verify the nature of flare plasma we investigate ten M-class flares in view of multi-thermal index $\delta$.

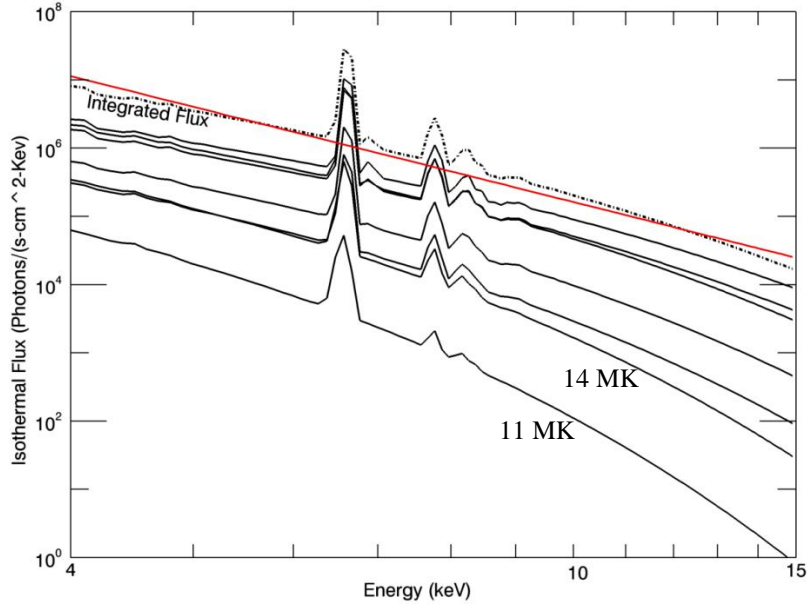

Figure 1: Theoretical line and continuum X-ray spectral evolutions of isothermal flux in the energy range 4-15 keV considering the temperature and emission measure from 'A. Awasthi, private communication'. The temperatures are in steps of 3 MK (thin curve) and their sum is shown as the dotted curve. The red curve is the best fit considering a power-law function, revealing a power-law index of γ=-4.65.

## 3. Observations, Analysis and Results

In order to probe the nature of observed solar flare plasma in view of multi-thermal or isothermal nature in comparison with the simulated multi-thermal plasma described in the previous section we study full disk integrated X-ray emission from ten M-class flares observed by SOXS mission. The "Solar X-Ray Spectrometer (SOXS)" was launched onboard the Indian GSAT-2 spacecraft in 2003 (Jain *et al.*, 2005). The SOXS employs Si and CZT solid-state detectors. The Si detector provides unprecedented high spectral resolution (0.7 keV) in the 4-25 keV energy range. The flare characteristics of the selected ten flares are given in Table I. The start and peak time, and peak flux of the flares are measured in 7-10 keV as observed by the Si detector of the SOXS payload. The data have been taken from the SOXS website at http://www.prl.res.in/~soxs-data.

Table 1
Characteristics of flare plasma

| | Observed parameters | | | | | | Calculated parameters | | |
|---|---|---|---|---|---|---|---|---|---|
| Date | Start time | Peak time | Peak intensity[1] | GOES class | DEM index[2] ($\delta$) | DEM ($10^{49}$ cm$^{-3}$keV$^{-1}$) | Peak temperature[3] (MK) | Loop length [$10^9$cm] | $n_e$ ($10^{10}$ cm$^{-3}$) | $t_c$ [sec] |
| 19-11-03 | 03:58 | 04:02 | 3.07 | M1.7 | 4.38 | 0.033 | 26.38 | 4.52 | 10.9 | 579.85 |
| 07-01-04 | B03:55 | 04:00 | 6.50 | M4.5 | 5.10 | 0.144 | 21.01 | 7.38 | 8.24 | 1195.59 |
| 19-01-04 | 04:35 | 04:54 | 2.45 | M1.0 | 5.37 | 0.119 | 25.64 | 20.30 | 2.99 | 2990.58 |
| 25-04-04 | 05:28 | 05:36 | 5.80 | M2.3 | 5.39 | 0.048 | 25.88 | 15.60 | 4.75 | 2753.94 |
| 14-08-04 | 04:13 | 04:14 | 8.80 | M2.4 | 5.32 | 0.185 | 21.63 | 20.20 | 5.28 | 1398.03 |
| 31-10-04 | 05:26 | 05:31 | 6.04 | M2.3 | 5.69 | 0.155 | 26.48 | 30.80 | 1.82 | 4640.42 |
| 03-06-05 | 04:06 | 04:10 | 1.70 | M1.3 | 5.15 | 0.043 | 16.19 | 7.79 | 2.91 | 295.89 |
| 03-08-05 | 04:57 | 05:05 | 5.40 | M3.4 | 5.31 | 0.119 | 22.25 | 20.90 | 1.77 | 1817.76 |
| 12-09-05 | 04:31 | 05:00 | 1.48 | M1.5 | 5.11 | 0.079 | 19.13 | 8.34 | 5.88 | 1099.12 |
| 17-09-05 | 06:01 | 06:05 | 14.5 | M9.8 | 5.39 | 0.092 | 20.93 | 4.34 | 29.3 | 1676.81 |

1. The X-ray peak intensity is in units of $10^4$ photons cm$^{-2}$ s$^{-1}$ keV$^{-1}$ in the 7-10 keV band.
2. $\delta$ is the power law index for DEM dependence on $T$.
3. Average of the peak temperatures over the temporal evolution of the flare.

### 3.1 Spectra Analysis:

We use the OSPEX (Object Spectral Executive) software package inside SolarSoft to analyze the data. The OSPEX is an object-oriented interface for X-ray spectral analysis of solar data. It is a new version of the original SPEX (Spectral Executive) code written by R. Schwartz in 1995. Through OSPEX, the user reads and displays the input data, selects and subtracts the background, selects time intervals of interest, selects a combination of photon flux model components to describe the data, and fits those components to the spectrum in each selected time interval. During the fitting process, the response matrix is used to convert the photon model to the model counts to compare it with the input count data. The resulting time-ordered fit parameters are stored and can be displayed and analyzed with OSPEX. The entire OSPEX session can be saved in the form of a script and the fit results are stored in the form of a FITS file. The OSPEX enables us to look at the temporal evolution, spectrogram and spectral evolution. It also enables us to fit energy spectra using the CHIANTI code (Dere *et al.*, 1997) for flare plasma diagnostics with the application of various thermal, line emission, multi-thermal, and non-thermal functions.

Thus to calculate the peak plasma temperature and the DEM we have employed the least-$\chi^2$ fitting to the SOXS X-ray spectra in OSPEX as described earlier by Jain *et al.* (2008).

However, we have discarded spectra with pulse pile-up to avoid spurious results. Figure 2 shows spectral evolution of the flare event of 19 November 2003 in the energy band of 4-25 keV. The spectral fit is made applying the CHIANTI code with the help of multi-thermal and single power-law functions provided in OSPEX. The modeled flux contribution from the multi-thermal function is shown by the green-colored line and that of the single power-law function with a break-energy point at 12 keV is shown by the yellow line. The total of these two contributions is shown by the red line. The total modeled flux is a good fit over the observed spectra with the goodness of fit evaluated as $\chi^2$=1.84. For this event, we estimate DEM=$0.1*10^{49}$ cm$^{-3}$ keV$^{-1}$, peak temperature $T$=24.4 MK and a power-law index of $\delta$=-4.77 from spectral fitting. The same method is employed for all flare events under consideration and the average of estimated values of DEM and $\delta$ is listed in Table 1. We find the power-law index in the range of -4.4 and -5.7. These values are in good agreement with the expected value of $\delta$= -5.15 (*cf.* Section 2) obtained by the simulation and thereby verify the nature of flare plasma to be multi-thermal.

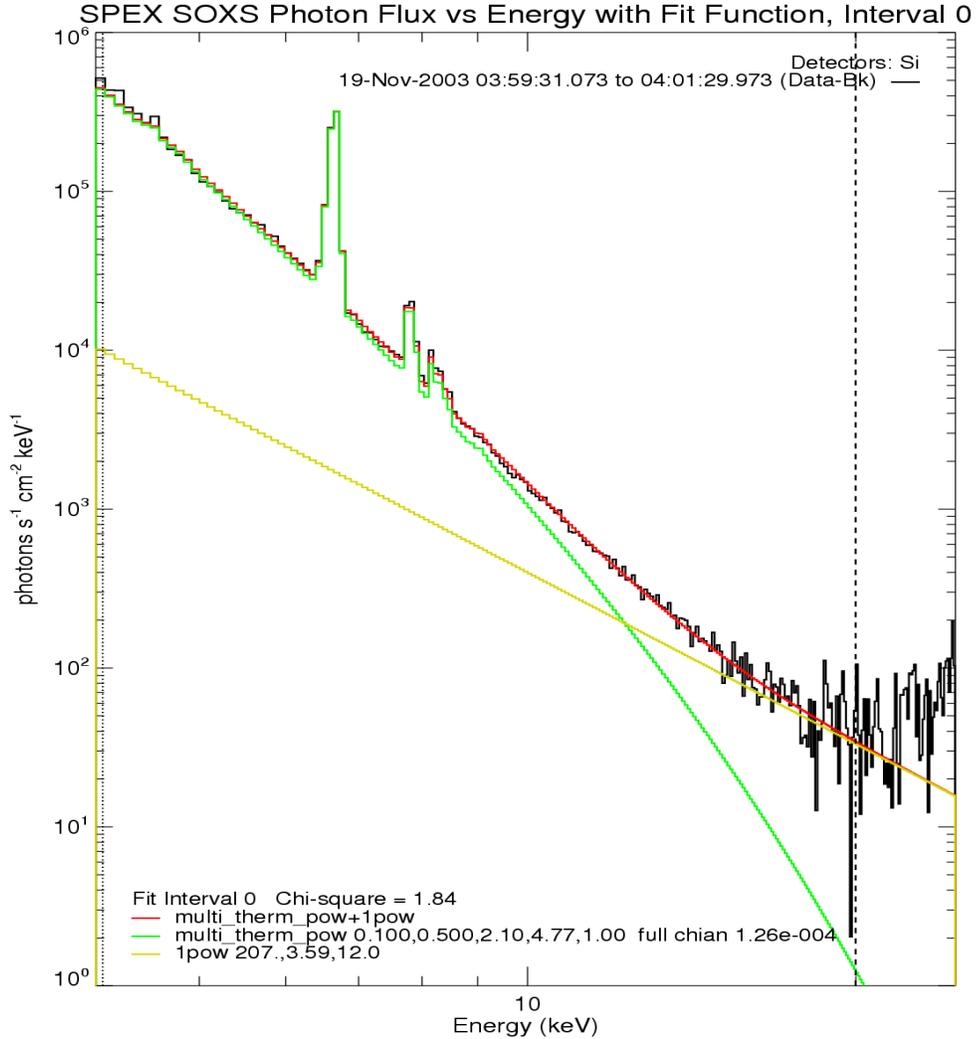

Figure 2: Spectral evolution of the 19 November 2003 flare event in the energy range of 4-25 keV as observed by the Si detector onboard the SOXS mission (shown by black line). The spectral fit with the help of multi-thermal, single power-law, and pulse pile-up functions provided in the OSPEX package is shown respectively by green and yellow lines. The goodness of fit of the total modeled flux (red line) over the observed spectra is evaluated to be $\chi^2$=1.84.

**3.2 Temporal Analysis:**

Figure 3 shows the temporal evolution of the 19 November 2003 flare in 13 energy bands. This flare evolved in the beginning via slow plasma heating seen in soft X-rays (4-10 keV), and then through impulsive transport followed by a long-duration thermal component. This figure also reveals a delay of the flare onset at higher energies, peaking earlier with respect to the low-energy curves. For example, Figure 3 shows that the flux in the highest energy band (20-25 keV) peaked at 03:59:14 UT, while it peaked in the next lower energy band (15-20 keV) at 03:59:30 UT, at a delay of 16 s. Figure 4 shows these delays as a function of energy, and demonstrates that the rise time *i.e.*, the time from the onset to the peak emissions, is longer at lower energies. In fact, this explains why the soft X-ray flare is of longer duration than the hard X-ray flare. However, this behavior can be explained in terms of the conduction cooling time. The electron density ($n_e$) calculated for all flares ranges between 1.77 and 29.3 * $10^{10}$ cm$^{-3}$ (*cf*. Table 1) suggests that the conduction cooling of thermal X-ray plasma dominates over radiative cooling in the initial phase of the flare (Culhane *et al.,*1970). The conduction cooling time is expected to become shorter with higher temperatures ($\tau_c \propto T^{-5/2}$) as thermal bremsstrahlung involves increasing photon energy with higher flare plasma temperature. Thus longer cooling delays are expected at low energies. Our findings from the analysis of ten flares confirm this scaling law. This suggests the multi-thermal plasma for all analyzed flares, and thereby the soft X-ray peaks are always delayed with respect to the hard X-ray peak. The cause behind the delay essentially could be due to the time delay between chromospheric heating, evaporation, and cooling of flare loops. Thus we may conclude that the multi-temperature components in the flare plasma are due to heating and cooling processes during the flare evolution.

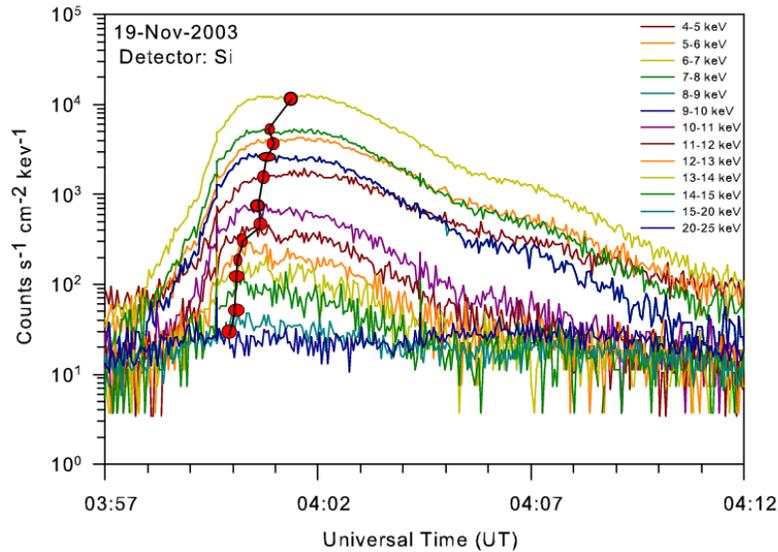

Figure 3: Temporal profiles in 13 energy bands ($\varepsilon$ = 4-5, 5-6 …14-15, 15-20, and 20-25 keV) of the 19 November 2003 flare. Each peak time is shown by circle. The curve connecting these circles illustrates the delay in peak time between successive profiles.

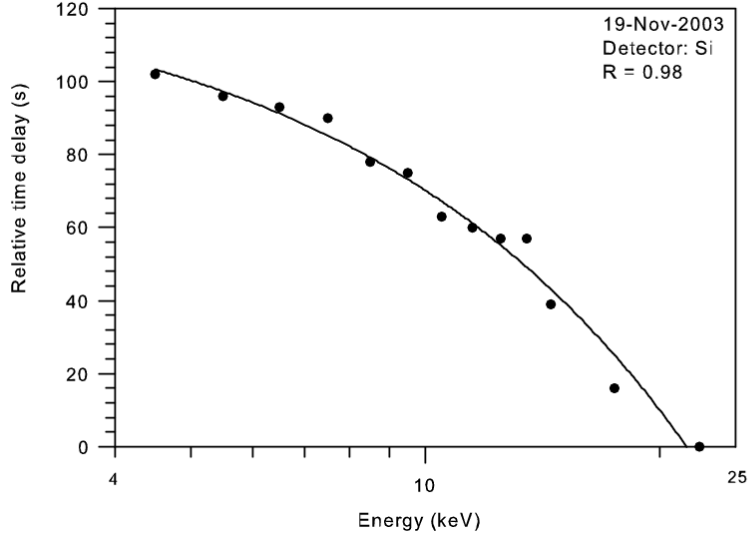

Figure 4: Relative time delay of the peak in a given energy band with respect to the reference energy band (20-25 keV) as a function of energy for the 19 November 2003 solar flare. The best-fit $\chi^2$ value is 0.98.

On the other hand, Hudson (1991) and Dennis and Zarro (1993) found that the thermal emission often closely resembles the integral of the non-thermal emission, which is known as the Neupert effect. However, it is difficult to know precisely the break-energy ($E_B$) where thermal contribution ends and non-thermal emission begins, and as well as the fraction of thermal and non-thermal contributions in the total X-ray energy budget (Jain *et al.*, 2005). In the current investigation, therefore, we attempt to measure the thermal-nonthermal crossover energy *i.e.*, the break-energy ($E_B$) point with the help of the behavior of the relative time delay ($\Delta t$) and the conduction cooling time ($\tau_c$). Our time domain technique is new and simpler in contrast to finding the break energy with the help of spectral evolution of flare plasma emission.

The cooling at a given energy band can be approximated from the time profile of the flare. However, flare plasma cooling at initial high temperatures ($T \geq 10$ MK) is dominated by conduction cooling and is observable in soft X-rays, while the later phase detected in EUV ($T \leq 2$ MK) is dominated by radiative cooling. The conduction cooling time ($\tau_c$) is given by the ratio of the thermal energy to the conductive loss rate, and can be expressed as a power-law dependence on the temperature

$$\tau_c(T) = \tau_{c0} \left(\frac{T}{T_0}\right)^{-\beta}, \qquad (3)$$

where, $T$ is the flare peak temperature, $T_0 = 11.6$ MK and $\beta = 5/2$, and the thermal conduction cooling time scale $\tau_{c0}$ is given by

$$\tau_{c0} = 344 \left(\frac{L}{10^9}\right)^2 \left(\frac{n_e}{10^{11}}\right), \quad (4)$$

where $L$ is the half loop length and $n_e$ is the electron density and are estimated as follows.

We have estimated the length and the volume of the flaring loops with the help of 171 and 195 Å images from TRACE and EIT observations and Hα images recorded at the Hiraiso Solar Observatory (HSO) in Japan (Akioka, 1998). The EIT and TRACE observations are downloaded from the URL http://idc-solar.ias.u-psud.fr/ and Hα full-disk images of HSO are obtained from URL http://sunbase.nict.go.jp/solar/db/home.html/. We process the data with the help of routines provided in the SolarSoft package in IDL. Firstly, we select the active region in which the flare occurred. This selected region is substantially magnified in order to identify and extract the loop structure. The length of the identified loop structure is measured using a pixel difference method. In this technique the array of pixels is manually selected that cover the loop structure. The pixel array on the spatial image has coordinates *viz.* $[(x_1, y_1), (x_2, y_2), (x_3, y_3),...]$ in the selected region. The loop length of the identified loop structure is then measured as

$$\text{Loop length } 2L = \sum_{i=1}^{n} \sqrt{(x_i - x_{i-1})^2 + (y_i - y_{i-1})^2} \text{ (in pixel)}, \quad (5)$$

where the loop length $2L$ is in pixels and is converted to arcsec with the help of the pixel scale of the respective instruments. The pixel scales of TRACE, EIT and HSO are respectively, 0.4993, 2.6 and 1.2428 arcsec. The quantity *n* is the dimension of the pixel array. However, in order to estimate the values of $\tau_{c0}$ and $\tau_c$ we employ the half loop length ($L$), and volume $V$ estimated by a cylindrical approximation as follows.

$$V = \pi r^2 (2L) \text{ (in cm}^3\text{)} \quad . \quad (6)$$

Here, $r$ is the radius of the loop and estimated by the half of the loop diameter $D$. $D$ is estimated by averaging the observations of the loop cross-section at three different places $d_1$, $d_2$, $d_3$ (*cf.* Figure 5: top left panel) on the loop. The loop length $2L$ has been estimated from the coronal loop images obtained in either of the instruments TRACE (171 Å), EIT (195 Å) or Hα at the peak emission period. For example, shown in Figure 5 are the images of the coronal loop emission for the flare event of 19 November 2003 observed in 171 Å by TRACE at 04:28:59 UT, in 195 Å by EIT at 04:00:10 UT and in Hα by HSO at 04:00:02 UT. The peak X-ray emission for the 19 November 2003 event observed by the Si detector is at 04:00:01 UT, and therefore, for this event, we have considered the EIT image to process for the loop length. Post-flare loops are visible in all wavelengths. The cross symbols on the images represent the identified loop structure. The loop-top and one of the foot-points are shown by A and B respectively.

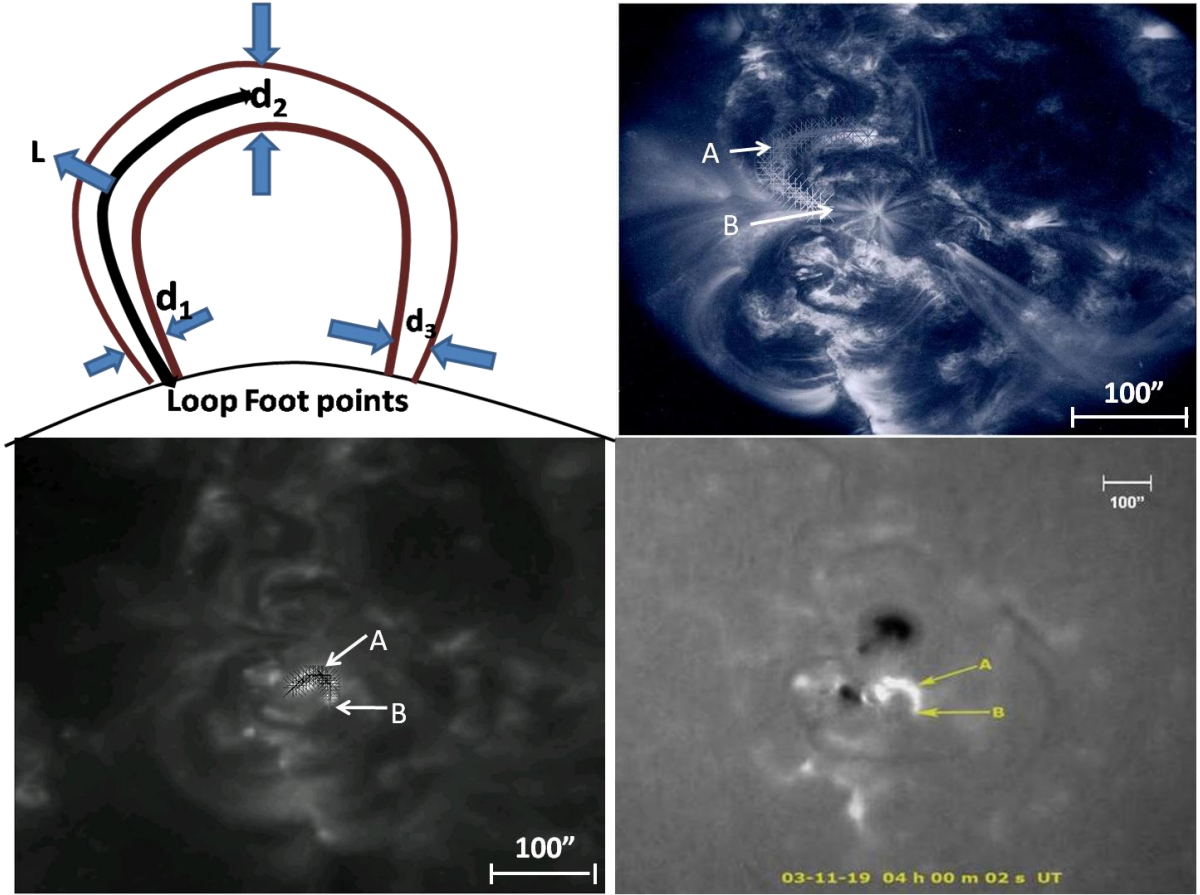

Figure 5: (Top left) The schematics of a length and diameter estimations of a typical flare loop. (Top right) Flare event of the 19 November 2003 observed in TRACE 171 Å. The loop is shown in white crosses. (Bottom) EIT and HSO observations of the same event in 195 Å and in Hα respectively. The loop-top and one of the foot-points are shown respectively by points A and B.

The estimated volume of a flare loop thus enables us to estimate the electron density $n_e$ with the help of DEM estimated from the X-ray spectra fitting (Equation (2)). The conduction cooling time as a function of energy is thus estimated with the help of Equations (3) and (7), and the relation between the peak temperature ($T$) and energy ($\varepsilon$) for which the multi-thermal bremsstrahlung spectrum $F(\varepsilon)$ contributes most (Aschwanden, 2007),

$$\varepsilon(T, \delta) = k_B T(\delta + 0.5). \tag{7}$$

The energy-dependent conduction cooling time is given as

$$\tau_c(\varepsilon) = \tau_{c0} \left[\frac{\varepsilon}{(\delta+0.5)}\right]^{-2.5}, \tag{8}$$

where ε is the energy and $\tau_{c0}$ is conduction cooling time scale obtained from Equation (4).

The X-ray time profiles of Gaussian nature, however, without cooling when convolved over the conduction cooling time allows us to compare with the observed time profiles as expressed as,

$$f(\varepsilon, t) = F_0(\varepsilon) \int_{-\infty}^{t} exp\left[-\frac{(t'-t_0)}{2\tau_g^2}\right] exp\left[-\frac{(t-t')}{\tau_c}\right] dt'. \tag{9}$$

Here, $t'$ is the time over which the conduction cooling has evolved and $t_0$ is the flare peak time. Therefore, the peak of the observed time profile actually becomes delayed with respect to the peak of the un-convolved Gaussian time profile. The Gaussian width ($\tau_g$) is obtained by fitting the light curve of 20-25 keV energy band with the gaussfit.pro procedure provided in IDL. The application of $L$ and $n_e$ in Equation (4) allows us to estimate the thermal conduction cooling time scale ($\tau_{c0}$) to be 1920 s and using Equation (8), we obtain $\tau_c$ to be 284 **s** at energy ε=12.5keV for the flare event of 19 November 2003. The analysis of all flare events under current investigation shows that $\tau_{c0}$ varies between 1520 and 18600 s, and averaged $\tau_c$ between 296 and 4640 s (*cf.* Table 1). Figure 6 shows the behavior of the observed time delay normalized with respect to $\tau_g$ ($\Delta t/\tau_g$) and the conduction cooling time normalized with respect to $\tau_g$ ($\tau_c/\tau_g$) for all events under consideration.

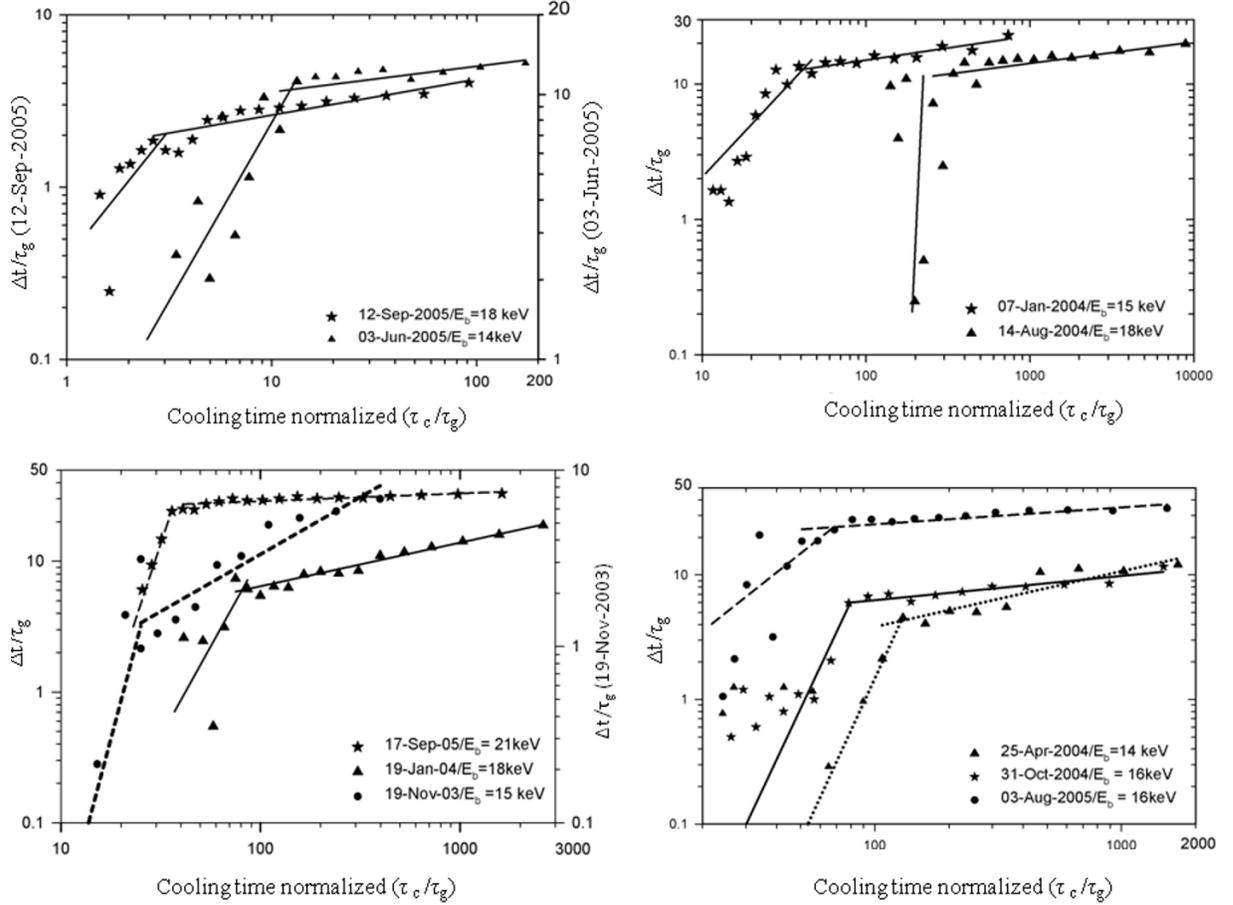

Figure 6: Observed time delay normalized with respect to $\tau_g$ ($\Delta t/\tau_g$) as a function of the conduction cooling time normalized with respect to $\tau_g$ ($\tau_c/\tau_g$) for the flare events.

Our observations and analysis as shown in Figure 6 reveal that for a relatively short cooling time, the delay $\Delta t/\tau_g$ vanishes because of the negligible involvement of conduction cooling time, while in the opposite limit the delay $\Delta t/\tau_g$ increases and saturates beyond certain energy in the asymptotic limit of large cooling time for a given flare event. This suggests that the X-ray emission beyond this energy involving longer conduction cooling time is thermal in nature and before this energy is of non-thermal nature. This technique enables us to estimate the thermal-non-thermal crossover energy (break energy) within the energy resolution limits of the Si detector (0.7 keV). We find the break-energy ($E_B$) point to vary between 14 and 21±1 keV for the flares considered in this investigation.

## 4. Discussion

The single temperature approximation has been widely used in the past to investigate X-ray emission from the solar flares. However, the fact that the plasma is heated at different temperatures (multi-thermal plasma), and therefore the emission measure varies as a function of the temperature, emphasizes the crucial need to study X-ray spectra with improved energy and temporal resolution (Aschwanden, 2007, 2008; Jain *et al.*, 2008). The present study of solar flares, in this context, using the Si detector of SOXS with the energy resolution <1 keV is a step ahead in this direction. In order to model the multi-temperature X-ray emission from the solar flare, firstly we begin with plasma parameters *viz:* $T$ and EM considered by 'A. Awasthi, private communication' to model isothermal plasma flux and then integrate the series of this isothermal plasma flux to obtain multi-temperature plasma flux. The integrated flare spectra of individual temperatures has a shape of a power-law and resembles very well the observed spectra, suggesting that the flare X-ray emission up to 25 keV is multi-thermal in nature. Aschwanden (2007) employed 10-50 keV X-ray spectra from RHESSI and fitted the 30-50 keV emission assuming non-thermal contribution, and extrapolated it into the low energy regime. He then subtracted the extrapolated part from the observed flux to deduce pure thermal contribution in the 10-30 keV energy band. However, we found in the current investigation insignificant contribution of thermal component above 20 keV. Thus we suggest that between 20 and 30 keV there must be significant non-thermal contribution. Further, we find the multi-thermal nature in all ten flares under investigation during their whole duration, which is in contrast to Phillips *et al.* (2005) who suggested the isothermal nature of flare plasma during decay phase.

Further, there are strong indications that, in many flares the non-thermal component contains a substantial fraction of the total energy (Jain *et al.,* 2000, 2005; Gan *et al.,* 2001; Lin *et al.,* 2002). The flare-accelerated 10-100 keV electrons appear to contain a significant fraction ≈ 10-50% of the total energy, indicating that particle acceleration and energy release processes are intimately linked. How the Sun releases this energy, presumably stored in the magnetic fields of the corona, and how it rapidly accelerates electrons and ions with such high efficiency, and to such high energies, is presently unknown. However, to address these questions, efforts using X-ray techniques have been undertaken with limited resolution. Bursts of bremsstrahlung hard X-rays (≥20 keV), emitted by accelerated electrons colliding with the ambient solar atmosphere, are the most common signature of the impulsive phase of solar flares. Provided that the electron energy $E_e$ is much greater than the average thermal energy, $kT$, of the ambient gas, essentially all of the electron energy will be lost to Coulomb collisions (Jain *et al.*, 2000). Therefore, the non-thermal hard X-ray fluxe observed in many flares indicates that the energy in the accelerated > 20 keV electrons must be comparable to the total flare radiative and mechanical output (Lin and Hudson, 1976). Thus, the acceleration of electrons to tens of keV may be the most direct consequence of the basic flare-energy release process. High-resolution spectroscopy is the key to understanding the electron acceleration and energy release processes in solar flares. Therefore the precise measurement of the low-energy cut-off of non-thermal emission is an important quantity: not only it is related to the acceleration mechanism, but it also determines the total number of accelerated electrons and the energy they carry (Gan *et al.*, 2001). Usually the low-energy cut-off is assumed to be 20 or 25 keV, which constitutes a main ingredient of the so-called standard picture of solar flares. However, the variation in the low-energy cut-off indicates that a higher low-energy cut-off means that the total energy carried by non-thermal electrons is much less than that is required to explain the flare energy budget. But the great decrease of the total number of accelerated electrons seems to explain the number problem in the standard flare scenario. Thus it would be proper that further studies based on observations with high energy resolution are being carried out (Gan *et al.*, 2001). The low energy cut-off of the non-thermal hard X-ray emission commonly taken to be 20 to 30 keV is in fact not known due to the presence of thermal emission, which in extreme cases can extend to energies >30 keV. In principle, this lower energy cut-off of power-law electrons ($E_c$) could be related to the break point energy ($E_B$). Gan *et al.* (2002) obtained the probability distribution of $E_c$, which represents that about half is larger than 45 keV and another half smaller than 45 keV. In order to estimate the break-energy ($E_B$) between thermal and non-thermal energy release in a solar flare we find that the peak of the observed flare gets delayed from the 20-25 keV Gaussian time profile. However, the delay Δt (*cf.* Figure 6) beyond certain energy (break-energy) in the asymptotic limit of cooling time saturates for a given flare event, suggesting that the X-ray emission beyond this energy is of non-thermal nature. This break-energy ($E_B$) point is measured within the energy resolution limits of the Si detector (0.7 keV) and found to vary between 14 and 21±1 keV for the flares considered in this investigation (*cf.* Figure 6) in agreement to the break energy point to be ≈18±3.4 keV previously obtained by Aschwanden (2007).

## 5. Conclusions

We analyzed the multi-thermal evolution in solar flares in the energy range of 4-25 keV of ten M-class flares using the high-resolution spectra from the Si detector of SOXS. We model the spectral-temporal evolution of the soft and medium-hard X-ray flux $F(\varepsilon,t)$ in terms of an evolving multi-temperature plasma and found that the resulting photon spectrum is a power-law function of energy. Our observations provide direct evidence of multi-thermal plasma in all ten flares and suggest that the flare X-ray spectra should be fitted with multi-thermal rather than isothermal model. The relative delay measurements show that the rise-to-peak time is longer for lower energy, which further indicates that flares are composed of multi-thermal plasma. We also found that $\Delta t$ saturates for an asymptotic limit of large cooling time between 14 and 21±1 keV for the flares considered in this investigation, which suggests that the break-energy point varies between these energy ranges, in agreement to other investigations.

## Acknowledgements

The authors express their sincere thanks to Prof. M. Akioka for his kind support to provide Hα observations, and to the IAS data server to obtain TRACE and EIT observations of the flares analyzed in this investigation. They also convey their sincere gratitude to an anonymous referee for his/her comments and useful suggestions to improve the paper.